\documentclass[letterpaper,twocolumn,10pt]{article}
\usepackage{usenix-2020-09}

\usepackage{graphicx}
\usepackage{soul,enumitem}
\usepackage[noabbrev,capitalize]{cleveref}
\usepackage{subcaption}

\newcommand{\zt}{\texttt{zns-tools}}
\newcommand{\ioc}{\texttt{ioctl()}}

\begin{document}

\date{}

\title{Understanding (Un)Written Contracts of NVMe ZNS Devices with \zt{}}

\author{
{\rm Nick Tehrany, Krijn Doekemeijer, and Animesh Trivedi}\\
VU, Amsterdam
} 

\maketitle

\begin{abstract}
Operational and performance characteristics of flash SSDs have long been associated with a set of Unwritten Contracts due to their hidden, complex internals and lack of control from the host software stack. These unwritten contracts govern how data should be stored, accessed, and garbage collected. The emergence of Zoned Namespace (ZNS) flash devices with their open and standardized interface allows us to write these unwritten contracts for the storage stack. However, even with a standardized storage-host interface, due to the lack of appropriate end-to-end operational data collection tools, the quantification and reasoning of such contracts remain a challenge. 
In this paper, we propose \texttt{zns.tools}, an open-source framework for end-to-end event and metadata collection, analysis, and visualization for the ZNS SSDs contract analysis. We showcase how \texttt{zns.tools} can be used to understand how the combination of RocksDB with the F2FS file system interacts with the underlying storage. Our tools are available openly at \url{https://github.com/stonet-research/zns-tools}. 
\end{abstract}

\section{Introduction}
The emergence of flash solid state drives (SSDs) has resulted in the redesigning of the host interfaces~\cite{2022-nvme-spec,2012-nvme-vs-ahci}, the block layer~\cite{2013-systor-mq,2019-atc-aync}, I/O schedulers~\cite{2019-atc-mqfq,2021-fast-d2fq}, file systems~\cite{2015-fast-f2fs}, applications~\cite{2021-FAST-rocksdb-flash}, and distributed systems~\cite{2012-nsdi-corfu}. 
Despite the aforementioned significant end-to-end changes to leverage the characteristics of flash storage, the reasoning and analysis of modern flash SSDs are governed by many \textit{``unwritten contracts''}~\cite{2017-eurosys-unwritten} regarding how user data should be stored, grouped, accessed, and deleted. In parts such expectations are considered unwritten as SSDs themselves are complex with many internal hidden details (flash chip sizes, layout, garbage collection (GC) units, buffer sizes)~\cite{2022-systor-ssd-study}. Furthermore, the complex architecture of the layered modern storage stack makes the end-to-end reasoning about the unwritten \mbox{contracts} challenging.

\begin{figure}[t!]
    \centering
    \includegraphics[width=.4\textwidth]{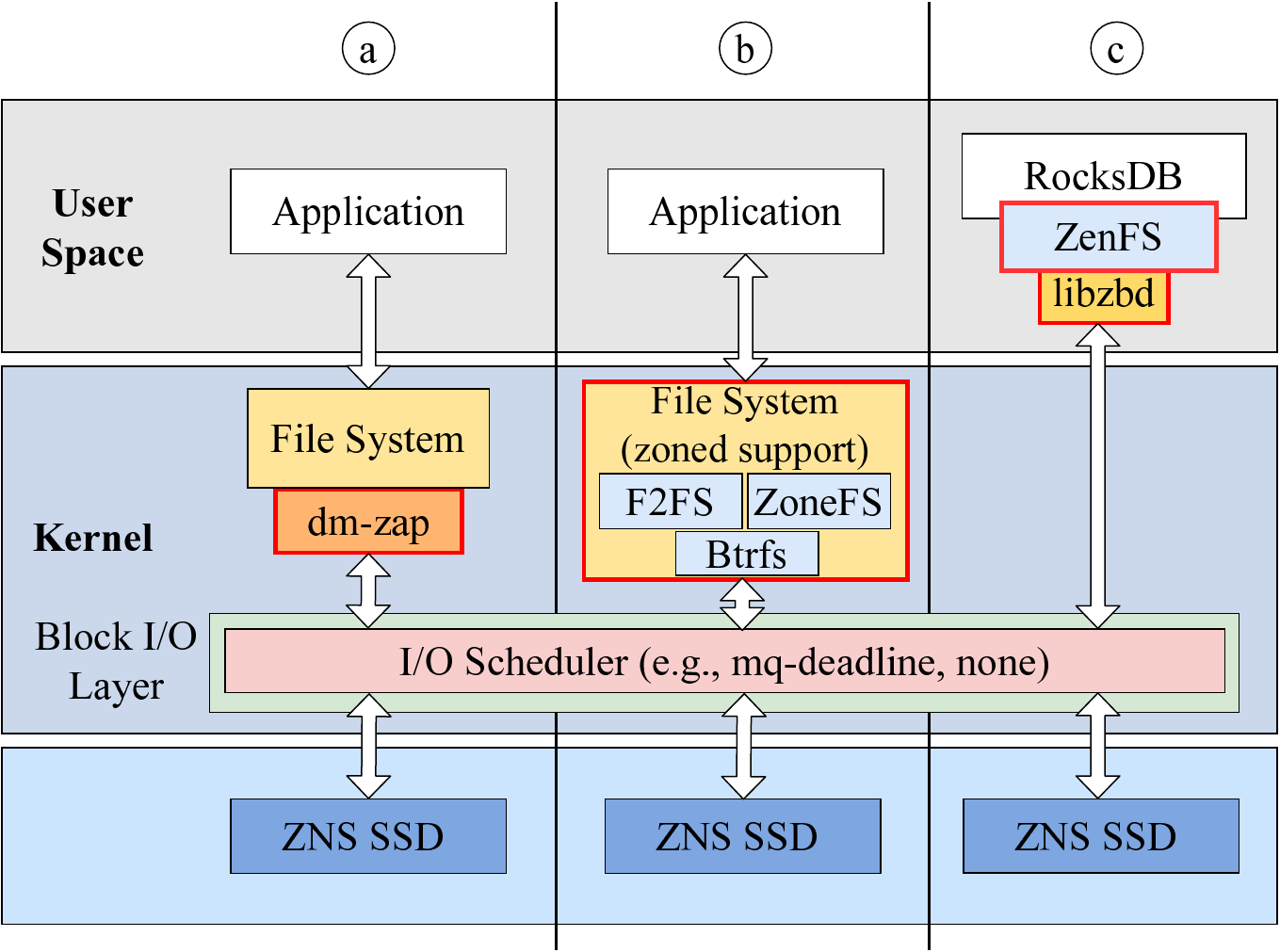}
    \captionof{figure}{ZNS integration at the (a) block layer; (b) file system level; and (c) application-level (example RocksDB). In all three cases, an end-to-end written contract analysis is missing.}
    \vspace{-0.3cm}
    \label{fig:zns-software-support}
\end{figure}

\textbf{NVMe Zone Namespace Devices and \st{Un}Written \mbox{Contracts}:} To address the challenges of complexity, researchers have advocated for more open interfaces~\cite{2017-fast-openchannel-ssd,2017-systor-autostream,2014-asplos-sdf}, which has culminated in the Zoned Namespace (ZNS) specification~\cite{2021-atc-zns,2021-wd-zns,2021-hotos-zone}. Briefly, a ZNS device divides its storage capacity into multiple \textit{zones} that can only be sequentially written (or appended to), thus closely mimicking how flash chips internally work. Data can be read from anywhere either sequentially or randomly. I/O on different zones is isolated and done in parallel, thus a zone represents a unit of parallelism. Beyond these basic read/write operations, ZNS devices have a number of unique flash/ZNS management-related commands. Of particular interest is the \textit{reset} command with zones, which explicitly resets a zone to be written again. In case there was any live data, it is the responsibility of the host software (file system, application) to ensure that the data is copied to a new zone before resetting the old zone. The significant advantage of such a design is that it clearly identifies the unit of garbage collection (zone) and the timing when to reset (under the control of the host software), both of these details are intricately complex to reverse engineer~\cite{2022-systor-ssd-study,2023-cluster-zns-study}. Hence, with ZNS devices, previously ``unwritten contracts'' are now standardized \textit{``written contracts''}, which forces us to re-think the operational design of our storage stacks~\cite{2021-hotos-zone}.

\textbf{Reasoning about the Written ZNS Contracts with Applications in a Layered Storage Stack:} ZNS software support in the storage stack is still under development. There are multiple ways through which ZNS contracts can be extended to an application in an end-to-end manner. Figure~\ref{fig:zns-software-support} shows such possibilities where ZNS software support is needed at the (a) block-level, I/O scheduler support with mq-deadline~\cite{2023-zns-sched}; or (b) file system level, currently F2FS and Btrfs support ZNS~\cite{2020-btrfs-zns}; or (c) application-level direct support, e.g., ZenFS customized file system backend for RocksDB~\cite{2021-atc-zns}. In such a layered architecture, it is challenging to ensure and cross-check if the ZNS contracts are extended and respected by all layers due to the \textit{semantic gap} between the layers. For example, the POSIX \texttt{fadvise} and \texttt{fctnl} calls are \textit{hints} to a file system that the file system is free to ignore. Similarly, 
 \mbox{ZNS}/flash file systems themselves have complex bookkeeping operations (garbage collection, log writing, fragmentation) where previously respected hints can be transparently overwritten by a file system without the application's knowledge (see below two examples in contract violation examples). Hence, in this work, we argue, \textit{there is a need to systematically investigate if and how the new written ZNS contracts are extended to applications}.


\begin{figure*}[t!]
    \centering
    \includegraphics[width=.8\textwidth]{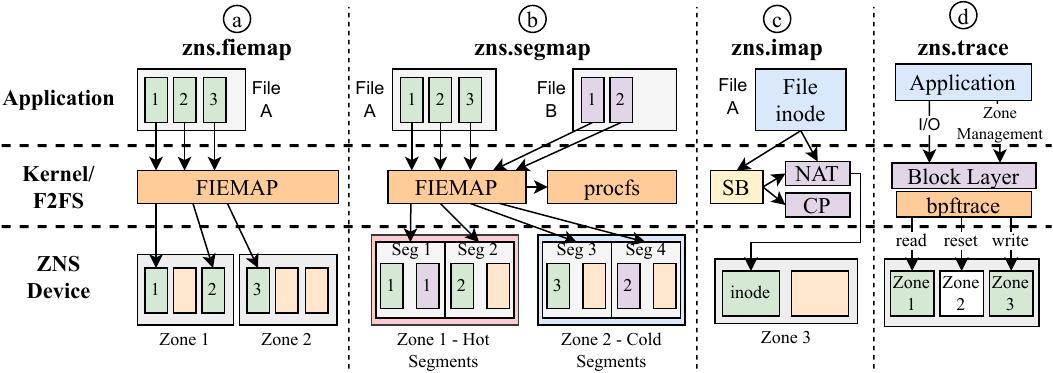}
    \captionof{figure}{Visual representation of \texttt{zns-tools} with (a) \texttt{zns.fiemap} mapping individual files to zones; (b) \texttt{zns.segmap} mapping multiple files to zones , segments and their lifetime classifications; (c) the \texttt{zns.imap} tool mapping the inode of a file to its ZNS zones (d) \texttt{zns.trace} tracing I/O and zone management activity to the ZNS zones.}
    \label{fig:zns-tools}
\end{figure*}

\textbf{\zt{}: A Framework to Study Written ZNS Contracts:} In this work, we propose a set of open-sourced tools called \zt{} for the ZNS software stack. Put together, these tools allow us to collect operational data and events across the stack in a programmed manner. The collected data is then analyzed and visualized to identify various contract violations. 
More specifically, we are interested in understanding the following previously ``unwritten contracts'' and how they are (not) followed across the layers: (1) ``Request Scale'' with ``Locality'': how data placement decisions are made (that result in large I/O, and parallel requests with minimal FTL overheads in ZNS SSDs) and how such allocations look on ZNS SSDs; (2) ``Grouping by death time'': how data grouping decisions are made to ensure minimal overheads from GC-related data movements and overheads; and (3) ``Uniform Data Lifetime'': how a ZNS device's lifetime is managed by generating and deleting data with a uniform lifetime, and subsequently explicitly resetting a zone in a ZNS SSD that (indirectly) controls wear-leveling on the device.

\textbf{Example: ZNS Contract violation with RocksDB and F2FS:}
\label{sec:introduction-contract-violation}
Using \zt{}, we identify two such violations about data grouping in the state-of-the-practice combination of RocksDB on F2FS with a ZNS SSD. 
The first issue is F2FS-specific. F2FS reclassifies the hotness (hot, warm, cold) of a data block after a round of GC cycle, where the GC always moves data blocks to cold data. This reclassification allows previously segregated data blocks under different hotness classes to be co-located in a new segment together, thus violating hotness hints from RocksDB. 
The second issue is RocksDB and F2FS specific. RocksDB writes SSTable to a file system in two passes, raw data (the table) and a small footer (less than a page). 
After writing the raw SSTable data to hot or cold data, depending on the level of the SSTable , RocksDB synchronizes the file causing F2FS to flush the data to the ZNS device. Upon completion of the flush, RocksDB writes the footer of the SSTable, containing a checksum, which is synchronized to the storage. Due to only a single page for the footer being dirty in the page cache, which is below the minimum threshold for F2FS of 16 pages, F2FS sets an inode flag to indicate the data as being hot. As a result, F2FS writes the footer page to the hot data segment, violating the contract for the SSTables by ignoring any hints for the SSTable writes.
Both of these incidents are examples of violating the ``Grouping by death time'' contract.

In this work we propose a collection of \zt{} that combined constitute a framework that allows parts-by-parts building an end-to-end understanding of how data is stored and managed in a layered storage stack on top of ZNS devices. The primary goal of the zns-tools framework is to allow building a complete picture of events across the layers that influence data placement and movement decisions. \cref{fig:rocksdb-timeline} shows on such end-to-end example generated from our framework.  Our key contributions in this work include the following: 
\begin{itemize}
    \item Making a case for building end-to-end operational data analysis tools to reason about previously unwritten, but now written ZNS SSDs storage contracts for applications. 
    \item \zt{}, an end-to-end framework that provides support for collecting, analyzing, and visualizing ZNS contracts across the layered storage stack. 
    \item Specific demonstration of the framework's capability on F2FS with RocksDB applications for an end-to-end visualization of operational analytics. \cref{fig:rocksdb-timeline} shows one such end-to-end example generated from our framework.   
    \item \texttt{zns-tools} are open-sourced and available here: \url{https://github.com/stonet-research/zns-tools}. This paper is available under CC-BY 4.0 License.
\end{itemize}

\section{Design of \zt{} Framework}
\zt{} is a new framework for ZNS SSDs that is designed to collect operational data about the ZNS storage stack to identify contract violations. It comes with tools that aid in extracting and visualizing: (\textbf{C1}) where is user data stored on ZNS SSDs; (\textbf{C2}) how much I/O is triggered to each zone; (\textbf{C3}) which zones are subjected to GC resets from the software stack and how data migrates across zones because of GC. Currently, there are four tools in the framework, \texttt{zns.fiemap}, \texttt{zns.segmap}, \texttt{zns.imap}, and \texttt{zns.trace}.
They are visualized in \cref{fig:zns-tools}, and their exact bash command and outputs are shown in our GitHub repository. 


\subsection{zns.fiemap}
\texttt{zns.fiemap} (ZNS file-mapping) is a tool that is designed to extract placement information from the stack to identify how the ``Locality'' contract is followed. The placement of files and data within modern SSDs is not static, and constantly changes in response to application or file system level events. 
Applications can typically provide allocation/location hints (by means of data temperature), but it is ultimately the file system that decides about the final data storage location. ZNS SSDs require any live data to be copied during GC, hence, the data location changes. Furthermore, any log-structured file system has its own out-of-place update mechanisms. With ZNS SSDs and log-structured file system, the location of the data is thus dynamic and changes constantly. The repeated execution of \texttt{zns.fiemap} traces the file data movement in this dynamic environment to identify how data from different files are grouped in zones. 
Furthermore, while moving file extents, if the ``Locality'' rule is not followed, that leads to file fragmentation with holes that are known to cause severe performance degradation~\cite{2021-Park-FragPicker,f2fs_defrag_tool,2020-date-ars,2018-Kadekodi-Geriatrix,2016-hotstorage-fs-frag-mobile}. Holes violate the ``Request Scale'' contract that recommends large sequential I/O requests. In ZNS, how a file is stored among zones also determines the amount of chip-level parallelism, a large file I/O can extract. Hence, \texttt{zns.fiemap} provides detailed information about the on-device zone-level placement of files within various file systems (addressing \texttt{C1},\texttt{C3}).

\texttt{zns.fiemap} retrieves file location mappings from the Linux kernel using the \ioc{} syscall with the \texttt{FIEMAP} flag. 
Support for \texttt{FIEMAP} is not necessary for file systems for POSIX compliance, however, all currently available file systems with ZNS support (F2FS and Btrfs) support this flag. With \texttt{FIEMAP}, file systems implement the tracking of extents, representing ranges of physically contiguous data for a file, which are returned to the \ioc{} caller. By iterating over the logical range of a file, \texttt{zns.fiemap} retrieves data mappings of all the extents for the particular file. 
The collected extents for the targeted file are mapped to their respective zone(s) containing the file's data using their logical address ranges. For example, on a ZNS SSD with a zone size of 1MiB, logical addresses between [0, 1MiB) fall within the first zone, [1MiB, 2MiB) on the second zone, and so on. The zone size and zone size ranges can be queried with the ZNS device using a zone management command. With all information about file extents, their addresses, address-to-zone mappings, \texttt{zns.fiemap} reports a detailed profile of the extent distribution (min, max, percentiles), hole statistics, and zone-level placement information. 
\cref{fig:zns-tools} (a) illustrates the operational aspect of \texttt{zns.fiemap}, retrieving the extent mappings of File \texttt{A} using \ioc{} with the \texttt{FIEMAP} flag, followed by mapping the three file extents to zones 1 and 2.

\subsection{\texttt{zns.segmap} and \texttt{zns.imap}}
\texttt{zns.segmap} and \texttt{zns.imap} are tools that collect and quantify metadata around the ``Grouping by death time'' contract. This contract is implemented within a file system that uses application-level lifetime hints (\texttt{RWH\_WRITE\_LIFE\_*} flags with the \texttt{fcntl} call), or its own data segregation policies based on data access heatmaps. It recommends that any data that is about to get deleted, or over-written, should be stored within a single GC unit. With this design, once data is invalidated, the GC unit can be simply reset without having to copy any live data, thus reducing GC overheads. Naturally, an accurate grouping requires coordinated efforts from the application, as well as the file system, over a lifetime of the data/file(s). Furthermore, such groupings should be constantly evaluated due to the dynamic nature of data placement with GC events within the file system.

\texttt{zns.segmap} and \texttt{zns.imap} are tools to extract grouping information for file data and metadata for F2FS. F2FS does heat-based data grouping in \textit{segments}. A file can have its data stored in multiple segments, and a single segment can contain data from multiple files. By default, F2FS uses 3 classes of classification (hot, warm, cold) on two types of data, file data, and file metadata (inodes). \texttt{zns.segmap} extracts the file-to-segment mappings using \texttt{zns.fiemap}, and reads the segment hotness classification from \texttt{procfs}\footnote{\url{/proc/fs/f2fs/nvme0n1/segment\_info}}. 
To locate the inode in F2FS, \texttt{zns.imap} firstly reads the F2FS superblock, which is written at a particular offset within the storage device, followed by parsing the superblock to identify the location of the node address table (NAT), where F2FS stores the block addresses of inodes, and the last checkpoint. Traversing the NAT, \texttt{zns.imap} retrieves the block address for the file inode to look up, followed by issuing a single read request to retrieve the inode. The retrieved inode is mapped to the zone in which it is contained. 
With these tools, we report for each F2FS segment, its hotness classification, number of file extents, the segment-to-zone mapping, and the inode-to-zone mapping. With this information, we can stitch a complete timeline of user data as it ages in the storage system (\cref{fig:rocksdb-timeline}). Though these tools are F2FS-specific, they can be extended to other file systems that do active data segregation. With F2FS, this information is available in the \texttt{/proc} file system\footnote{We also tried Brtfs with ZNS, but it was unstable.}. 
The file system-related metadata is also available as a part of low-level libraries (such as libext2fs or libf2fs) or can be generated~\cite{210526}.  
\cref{fig:zns-tools} (b) and (c) illustrate the design of \texttt{zns.segmap} and \texttt{zns.imap} tools. They retrieve extents for various files (\ioc{} call) and inodes (FS metadata walks) from F2FS, and retrieve F2FS segment lifetime classifications and the inode location.

\subsection{\texttt{zns.trace}}
\begin{figure}[t!]
  \centering
  \begin{subfigure}[b]{0.25\textwidth}
     \includegraphics[width=.85\linewidth]{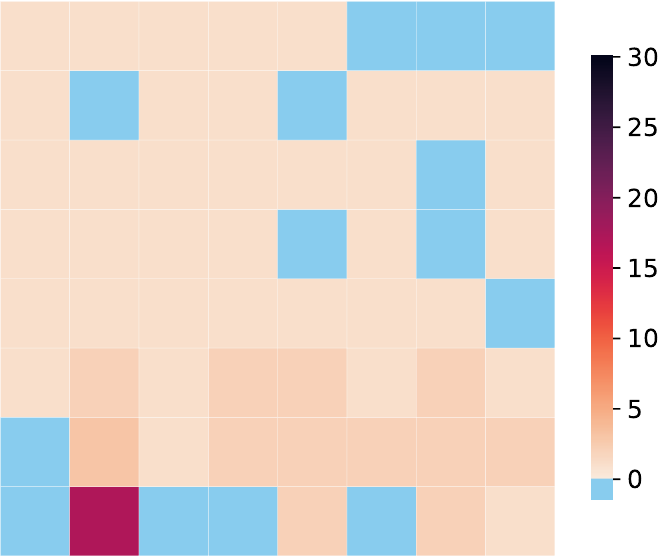}
    \caption{RocksDB + F2FS}
    \label{fig:heatmap-rocksdb-f2fs}
  \end{subfigure}%
  \centering
  \begin{subfigure}[b]{0.25\textwidth}
    \includegraphics[width=.85\linewidth]{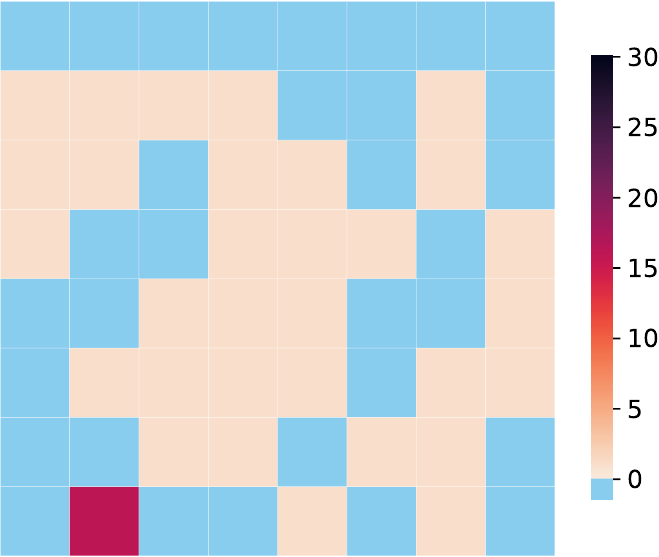}
    \caption{MongoDB + F2FS}
    \label{ig:heatmap-mongodb-f2fs}
  \end{subfigure}
  \centering
  \begin{subfigure}[b]{0.25\textwidth}
    \includegraphics[width=.85\linewidth]
    {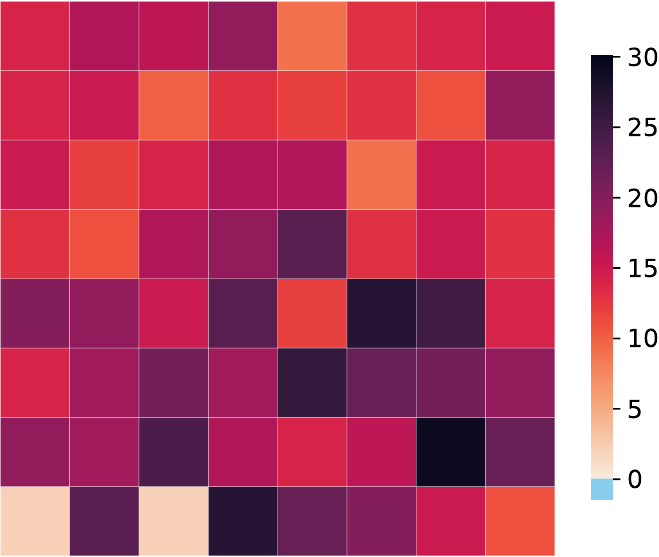}
    \caption{PostgreSQL + F2FS}
    \label{fig:heatmap-postgresql-f2fs}
  \end{subfigure}%
  \centering
   \begin{subfigure}[b]{0.25\textwidth}
     \includegraphics[width=.85\linewidth]{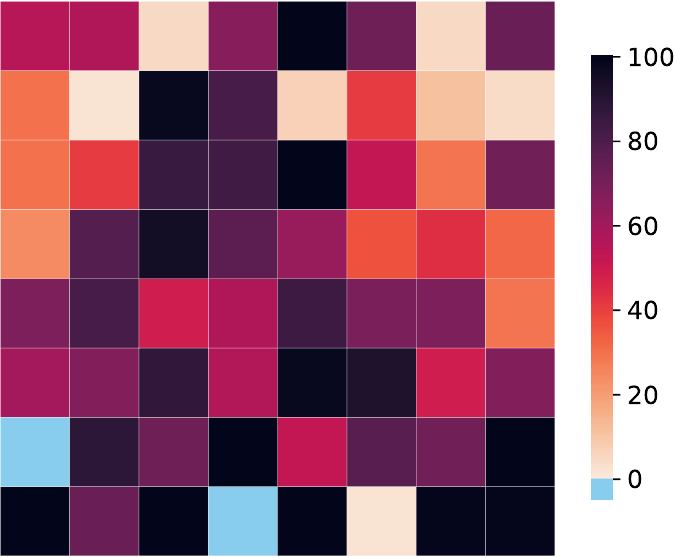}
    \caption{RocksDB + (aged) F2FS}
    \label{fig:heatmap-rocksdb-f2fs-aged}
  \end{subfigure}%
  \centering
     \caption{Reset visualization using zns.trace for an identical YCSB workload-A for multiple databases on F2FS. Figure (d) has a different heatmap scale (0-100).}
    \label{fig:heatmap}
    \vspace{-0.3cm}
\end{figure}

The \texttt{zns.trace} tool identifies performance-critical ``Request Scale'' and ``Uniform Data Lifetime'' contracts by tracking ZNS command calls. \texttt{zns.trace} is a combination of a BPFtrace~\cite{bpftrace} script and a plotting tool that traces zone-level ZNS write, append, read, and reset operations (addressing \texttt{C2}). Tracing such data is useful for investigating the access patterns to the underlying storage device independently of file system or application workloads.

\texttt{zns.trace} utilizes BPFtrace~\cite{bpftrace}, which inserts probes into the Linux kernel functions, that upon being triggered (i.e., the function being called) initiate data collection. The tracing script captures the NVMe I/O event with the command I/O sizes (their histogram) and zone reset calls, and maps the events to their corresponding zone(s). By analyzing the function arguments, the script identifies the type of operation, and further extracts data fields based on the request size. Importantly, the tracing supports ZNS devices in VMs (apart from just the host), where the zone reset command sets the function argument for the type of command to \texttt{REQ\_OP\_DRV\_OUT}, indicating the host driver (e.g., vfio-pci when using NVMe passthrough to the VM) is responsible for the request. This case requires to furthermore analyze the NVMe command and its zone management command field on the type of zone management action (e.g., close, finish, reset, open, offline). \texttt{zns.trace} relies on inserting kernel probes to parse I/O functions, and is therefore currently limited to kernel-based ZNS tracing. However, ongoing work is extending the tracing framework for user-level (e.g., SPDK) I/O and broader zone management request tracing. 



To illustrate the utility of \texttt{zns.trace} we run an identical workload on a number of database/file system and quantify the workload reset profiles. The number of resets is directly linked to the number of reset cycles that an SSD can undergo before exhausting flash chips. The ``Uniform Data Lifetime'' contract recommends generating data with a uniform lifetime to evenly spread the device wear. 
We run experiments on an emulated NVMe ZNS device (QEMU v6.0.0) with 64 zones of size 64MiB (4GiB in total). Our workload is a YCSB workload-A (50\% update, 50\% read)~\cite{2010-socc-ycsb} on RocksDB~\cite{rocksdb-git}(7.4.3), MongoDB~\cite{2018-springer-mongodb}(6.06), and PostgreSQL~\cite{momjian2001postgresql}(9.6.24) as KV backend targets with F2FS as file system. \cref{fig:heatmap} shows our results. Here each cell represents a zone, and the color indicates the total number of resets issued to the zone (since startup). Blue squares indicate a zone to which no reset commands were issued at all. There are two interesting observations. Firstly, for an identical workload, the three KV backends show vastly different profiles. 
We can see that for this test configuration PostgreSQL leads to significantly more resets. Additionally, we can identify that one of the bottom left zones (zone 2) is in all three cases heavily utilized. This zone corresponds to a warm node zone initialized by F2FS, where the inodes of files are written.
Secondly, file system aging has a significant impact on the reset profile as shown by \cref{fig:heatmap} (d) which has an aged F2FS (10$\times$ iterations). 
Both of these results make a case for a systematic study of how zone resets are called or consequently, how data grouping by uniform lifetime or death time is done.


\subsection{RocksDB on F2FS with ZNS Timeline}
\begin{figure}[t!]
    \centering
    \includegraphics[width=.45\textwidth]{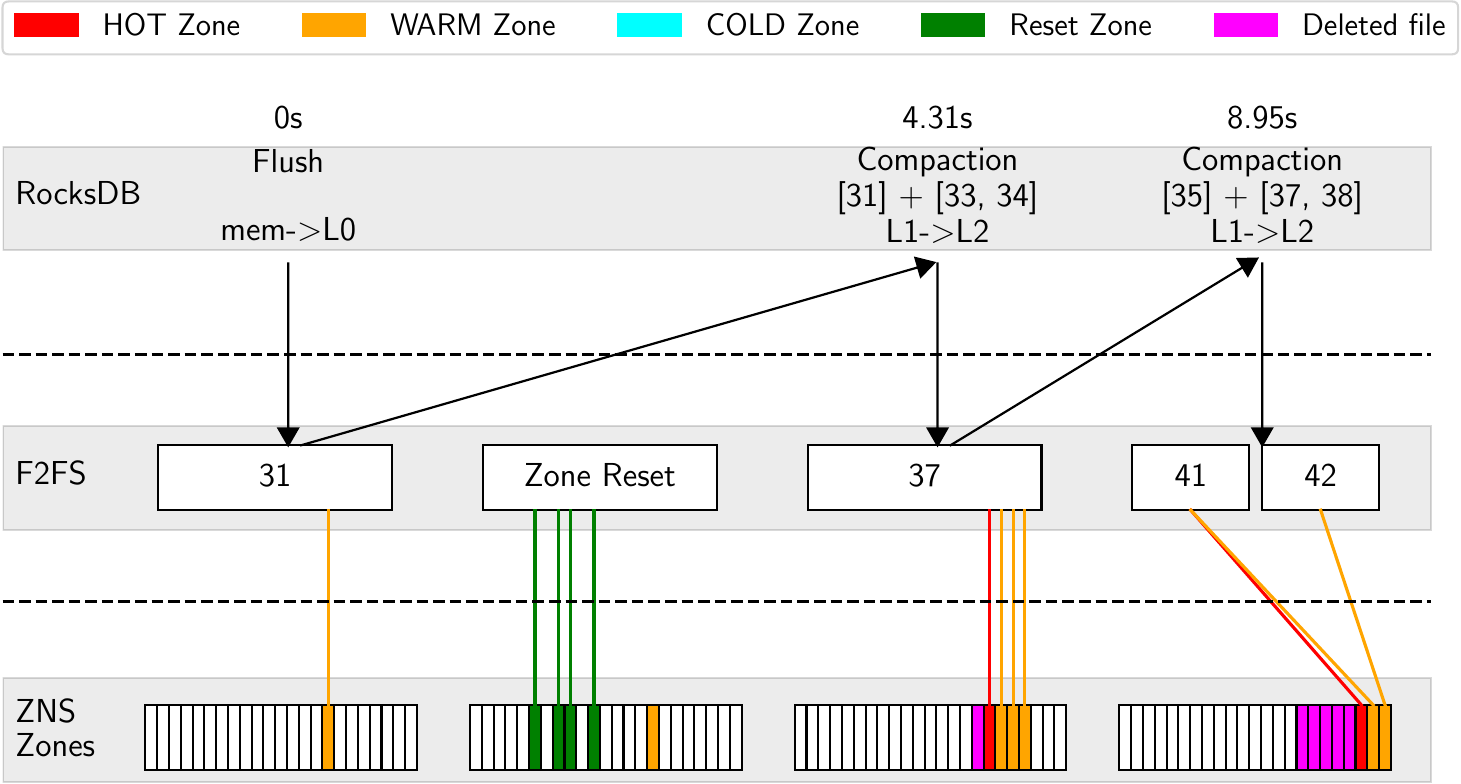}
    \captionof{figure}{(Abridged) End-to-end event timeline visualized using \zt{}.}
    \vspace{-0.3cm}
    \label{fig:rocksdb-timeline}
\end{figure}

To demonstrate an end-to-end utility of \zt{}, we build an end-to-end data placement visualization using RocksDB on F2FS. Using the RocksDB's db\_bench benchmark with workloads \textit{fillrandom} and \textit{overwrite} fills the F2FS file system on the ZNS device. Minor modifications (less than 100 LOCs) to the RocksDB source code are made to trace the activity of operations, and retrieve the data mapping of the generated files. On each compaction or flush operation, RocksDB calls the \zt{} to map all its files. 
\cref{fig:rocksdb-timeline} illustrates this end-to-end timeline (automatically generated from traces) with a few salient events across the layers (applications, file systems, and ZNS device) over the lifetime of an SSTable. The timeline starts at time 0, where data is flushed from memory to level-0 (abbreviated L0) as SSTable 31. The file is stored in the orange zone classified as WARM. Subsequently, F2FS issues a series of reset calls on several zones. The next logical event is the compaction of file 31 with 33 and 34 to generate file 37 on L2\footnote{internal trivial promotion events such as moving 31 from L0 to L1 are skipped in the visualization for the sake of clarity.}. While file 37 is written as WARM file, old files in prior zones are deleted (pink lines). In the next round, file 37 is picked up for compaction with 35 and 38 files to generate two L2 files (41, and 42). 
Such a timeline visualization gives an understanding of how the different RocksDB operation interaction with F2FS affects the utilization of the ZNS storage space, file classification, and data movement over time, thus making it easy to spot contract violations visually.

\section{Related Work}
Flash SSDs with their complex internal logic and \textit{``unwritten contracts''}~\cite{2017-eurosys-unwritten} have also been studied in detail for performance and operation characterizations~\cite{2021-tos-ssd-blackbox,2018-ieee-ssd-bottleneck-analysis,2010-mascots-ssd-blackbox-modeling,2011-mascots-ssd-modeling,2013-sigmetrics-ssd-expectations,2022-systor-ssd-study} and with the impact of GC operations~\cite{2018-tos-ibm-flashsystems,2019-tom-d-choice-gc,2021-sigmetric-ssd-management,2012-sigmetrics-ssd-perf-anamoly,2013-sigmetrics-gc-modeling}. 
There is a rich history of collecting file system traces, analyzing, and replaying them for understanding the impact of optimizations~\cite{2013-Jeong-AndroStep,1985-sosp-bsd-trace, 2003-lisa-nfs, 2000-atc-fs-comparison, 2004-fast-tracefs,DBLP:conf/fast/TarasovKMHPKZ12}. Much of these works only focus on basic read/write interfaces that are sufficient for HDDs, but not SSDs. 
None of these tools track device management-related operations (reset, open, finish, close), which are now a critical part of ZNS devices. The libzbd ZNS library also support a basic GUI visualization tool for zone states~\cite{2023-libzbd}, however, it lacks any application or file system level information. 
Beyond these tools, the eBPF-based BCC framework has emerged as the de-facto API for writing complex, end-to-end tracking frameworks~\cite{2023-bcc}, which we also leverage. 
Similar to \zt{}, IOScope uses eBPF-assisted file offset-based I/O tracing~\cite{2018-Saif-IOscope}. However, its tracing is limited to the files (at the VFS level), and does not connect the file to its location, which can change based on the file system and application level operations. Several block-level tracking tools exist (BCC's biotop and biosnoop, DTraces's iosnoop), however, they do not link block-level I/O back to the file system or applications. MapFS is a file system interface that presents low-level details from file to storage mappings for applications to manipulate data-heavy operations via light-weight metadata operations~\cite{2011-hotstorage-mapfs}. 
Prabhakaran et al.~\cite{2005-atc-stp} introduce techniques to study file system behavior with semantic knowledge of events and on-disk data structure layouts. \zt{} extends such motivation to include workloads with the new ZNS management operations as well.  
HintStore is a flexible framework that is designed to explore the effectiveness of hints with heterogeneous storage~\cite{2022-Ge-HintStor}. \zt{} analysis captures the after-effect of the hints, and its collected data can be used to verify if hints are implemented in an end-to-end manner. 
In a distributed setting, systems like Wintermute~\cite{2020-hpdc-wintermute}, Apollo~\cite{2021-hpdc-apollo}, and Beacon~\cite{2019-nsdi-io-supercomputer} provide a distributed framework for operational data collection and analysis. 
In comparison to these works, the focus of \zt{} is on collecting, analyzing, and visualizing written ZNS contracts across multiple storage stack layers to reason about previously unwritten SSD storage contracts.




\section{Conclusion and Ongoing Work}
In this work, we have presented a case to systematically reason about previously ``\textit{unwritten contracts}'' for flash SSDs on recently emerged ZNS flash SSD devices. Due to the unique I/O and management interface of ZNS SSDs, they make such unwritten contracts, written by putting them under the control of the host storage stack. To investigate the effectiveness of ZNS contracts, we have developed and presented a set of \zt{} to collect, analyze, and visualize ZNS operational data. 
We are working on tracing events from PostgreSQL, Aerospike, and scientific workloads on top of BrtFS also, to develop a broader framework in which such end-to-end contract analysis can be done.

\textbf{Acknowledgment:} This work in-parts is supported by the donations from Western Digital.

\section*{Availability}

\url{https://github.com/stonet-research/zns-tools}. 

\bibliographystyle{plain}
\bibliography{main}

\end{document}